\documentclass[12pt,twoside]{article}
\usepackage{fleqn,espcrc1}

\usepackage{epsfig} 
\newcommand{\AmS}{{\protect\the\textfont2
  A\kern-.1667em\lower.5ex\hbox{M}\kern-.125emS}}

\title{Nucleon structure effect on the longitudinal response function} 

\author{K. Saito\address{Tohoku College of Pharmacy, 
    Sendai 981-8558, Japan}}
       
\begin{document}

% typeset front matter
\maketitle

\begin{abstract}
Using the quark-meson coupling model, we study
the longitudinal response function for quasielastic electron
scattering from nuclear matter within relativistic RPA.  
In QMC the coupling constant between the 
scalar meson and the nucleon is expected to decrease with
increasing nuclear density.  
Furthermore, since the electromagnetic form factors of the in-medium nucleon
are modified at the same time, the longitudinal response function  
and the Coulomb sum are reduced by a total of about 20\% 
in comparison with the Hartree contribution.
\end{abstract}

\section{Introduction}

There is still considerable interest in the longitudinal response (LR) for
quasielastic electron scattering.  Within the framework of nonrelativistic
nuclear models and the impulse approximation,
it is very difficult to reproduce the observed, quenched LR \cite{exp}.  
In the mid '80s, several groups calculated
the LR function using the classic version of
Quantum Hadrodynamics (QHD-I) \cite{qhd}. It was argued that
the contribution of the relativistic random phase approximation (RRPA),
which includes vacuum polarization, is very important in reducing the
LR \cite{hor}.  

However, the nucleon has internal structure, and it is nowadays
expected that this structure should be modified in a nuclear 
environment.  In QHD-I \cite{qhd}
nuclear matter consists of {\em point-like\/}
nucleons interacting through the exchange of scalar
($\sigma$) and vector ($\omega$) mesons.  (More recent versions of
QHD \cite{newqhd} are capable of 
incorporating the effects of hadron internal structure.)    
Recently, we have developed a relativistic quark model for nuclear matter,
namely, the quark-meson coupling (QMC) model \cite{qmc}, which could be
viewed as an extension of QHD-I.  However, in QMC the mesons
couple to confined quarks and the 
nucleon is described by the MIT bag model.  Using QMC we here report  
the effect of in-medium changes in the structure of the nucleon on the
LR of nuclear matter \cite{lrf}.

\section{Relativistic RPA}

We first review the calculation of the LR
function for quasielastic electron scattering from (iso-symmetric)
nuclear matter in QHD-I.  The starting point
is the lowest order polarization insertion (PI), $\Pi_{\mu \nu}$, for the
$\omega$ meson.  This describes the coupling of a virtual vector meson
or photon to a particle-hole or nucleon-antinucleon
excitation: $\Pi_{\mu \nu}(q) = -ig_v^2 \int \frac{d^4k}{(2\pi)^4} 
\mbox{Tr}[G(k) \gamma_\mu G(k+q) \gamma_\nu]$, 
where $G(k)$ is the self-consistent nucleon propagator 
in relativistic Hartree approximation (RHA) \cite{qhd}.
We can separate the PI into two pieces: one is the density dependent part,
$\Pi_{\mu \nu}^D$, and the other is the
vacuum PI, $\Pi_{\mu \nu}^F$. 
The former is finite, but the latter is divergent and 
must be renormalized \cite{lrf}.  

In the Hartree approximation, where only the lowest one nucleon ring is
considered, the LR function, 
$S_L^H$, is simply proportional to 
$S_L^H(q) \propto G_{pE}^2(q) {\Im} \Pi_L(q)$.  
Here $\Pi_L (= \Pi_{33} - \Pi_{00})$ is the longitudinal (L) component of the
PI (we choose the direction of ${\vec q}$ as
the $z$-axis) and $G_{pE}$ is the proton electric form factor (EFF), 
which is usually parametrized by a dipole form in free space:
$G_{pE}(Q^2) = 1/(1 + Q^2/0.71)^2$ 
with the space-like momentum transfer, $Q^2 = - q_\mu^2$, in units of
GeV$^2$.  

In RRPA the L component of the PI, $\Pi_L^{RPA}$, involves the sum of 
the ring diagrams to all orders.
It involves $\sigma$-$\omega$ mixing in the nuclear medium, and
is given by  
$\Pi_L^{RPA}(q) = [(1 - \Delta_0 \Pi_s) \Pi_L + \Delta_0 \Pi_m^2]
/ {\epsilon_L}$, 
where $\epsilon_L$ is the L dielectric function \cite{lrf,soprop} and 
$\Delta_0$ is the free $\sigma$-meson propagator.  
Here $\Pi_s$ and $\Pi_m$ are respectively the scalar and the time
component of the mixed PIs.  
The vacuum component of the scalar PI is again divergent
and we need to renormalize it \cite{lrf,soprop}.  
For the mixed PI there is no vacuum polarization and
it vanishes at zero density.

\section{Nucleon structure effects}

To discuss the effect of changes in the
internal structure of the nucleon, 
we consider the following modifications to QHD-I:
\begin{enumerate}
\item Meson-nucleon (N) vertex form factor \\
Since both the mesons and nucleons are composite they have finite
size.  As the simplest example, we take a monopole form factor, $F_N(Q^2)$, 
at the vertex   
with a cut off parameter, $\Lambda_N = 1.5$ GeV \cite{lrf}. 
\item Modification of the proton electric form factor \\
We have studied the electromagnetic form 
factors of the nucleon in nuclear medium, using the QMC model \cite{emff}. 
The main result of that calculation is that the ratio of the EFF 
of the proton in medium to that in free space  
decreases essentially linearly as a function of $Q^2$, and that 
it is accurately parametrized as $R_{pE}(\rho_0,Q^2) \equiv
G_{pE}(\rho_0,Q^2)/G_{pE}(Q^2) \simeq
1 - 0.26 \times Q^2$ at $\rho_B = \rho_0$ ($=$ the normal nuclear matter 
density) \cite{lrf,emff}. 
\item Density dependence of the coupling constants \\
In QMC the confined quark in the nucleon couples to the $\sigma$ field
which gives rise to an attractive force.
As a result 
the coupling between the $\sigma$ and nucleon is expected to be
reduced at finite density \cite{qmc}. 
The coupling between the vector meson and nucleon 
remains constant because it is related to the baryon number. 
\end{enumerate}

To study the LR of nuclear matter, we first have
to solve the nuclear ground state within RHA \cite{lrf}.  
To take into account the modifications 1 and 3, we replace the
$\sigma$- and $\omega$-N coupling constants in QHD-I by: 
$g_s \to g_s(\rho_B) \times F_N(Q^2)$ and 
$g_v \to g_v \times F_N(Q^2)$, 
where the density dependence of $g_s(\rho_B)$ is given by solving the
nuclear matter problem self-consistently in QMC \cite{lrf}. 
Requiring the usual saturation condition for nuclear matter \cite{qmc}, 
we found the coupling constants:  
$g_s^2(0)=61.85$ and $g_v^2=62.61$ (notice that 
$g_s$ decreased by about 9\% at $\rho_0$).  
In the calculation we fix the quark mass, $m_q$, to be 5 or 300 MeV,  
$m_\sigma=550$ MeV  
and $m_\omega=783$ MeV, while the bag parameters are chosen so as
to reproduce the free nucleon mass with the 
bag radius $R_0=0.8$ fm \cite{qmc}.  
This yields the effective nucleon mass $M^*/M=0.81$ at $\rho_0$
and the incompressibility $K=281$ MeV.

\section{Numerical results}

\begin{figure}[htb]
\begin{minipage}[t]{80mm}
\psfig{file=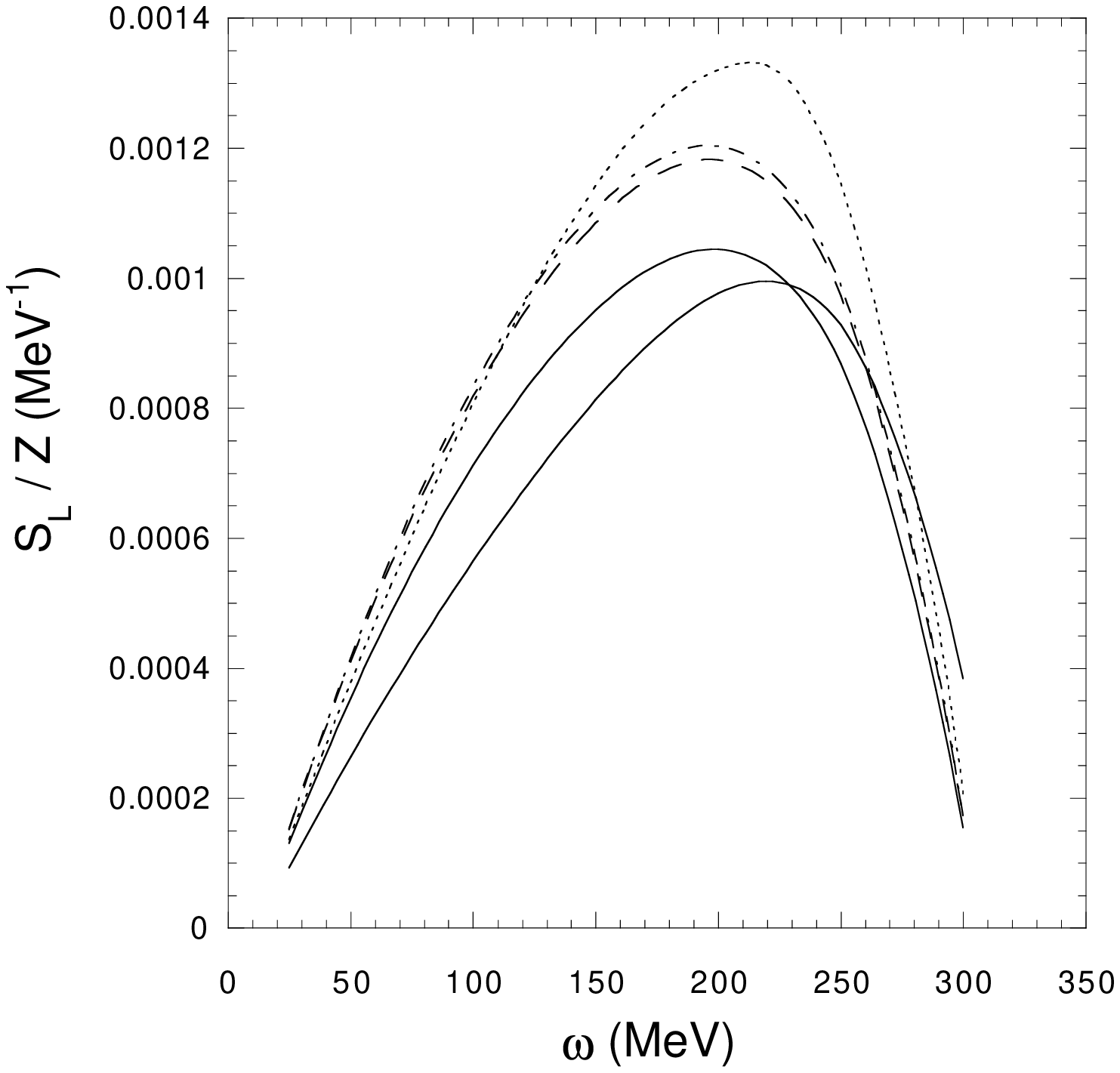,width=7.5cm} 
%\framebox[79mm]{\rule[-26mm]{0mm}{52mm}}
\caption{LR functions in QMC.  
We fix $q$ = 550 MeV and $\rho_B = \rho_0$. 
See text for details.  
}
\label{fig:resp}
\end{minipage}
\hspace{\fill}
\begin{minipage}[t]{75mm}
\psfig{file=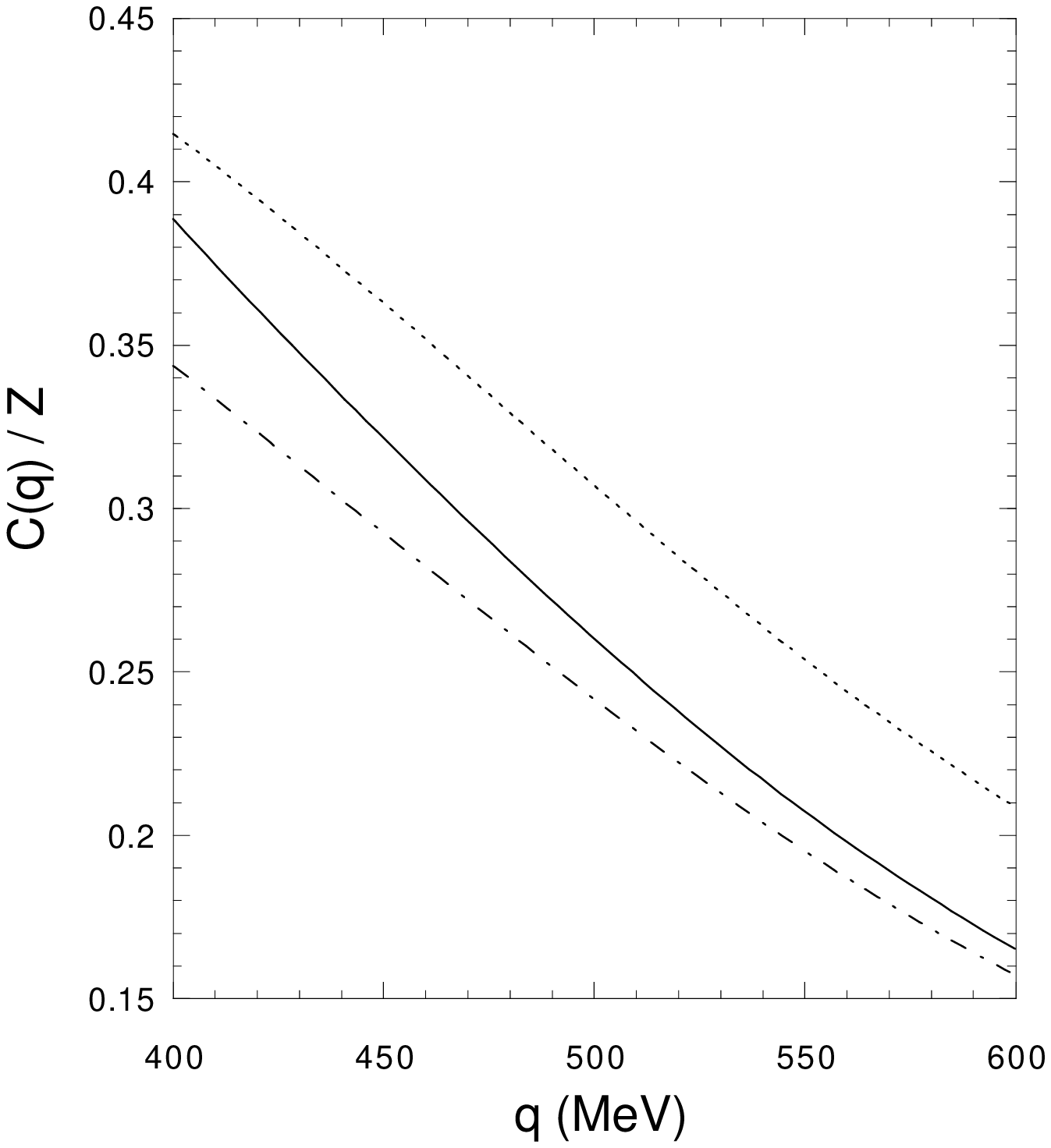,width=7.5cm}
%\framebox[74mm]{\rule[-26mm]{0mm}{52mm}}
\caption{Coulomb sum, $C(q)/Z$, at $\rho_0$.  
See text for details. 
}
\label{fig:csr}
\end{minipage}
\end{figure}

Our result is shown in Fig. \ref{fig:resp}.  In the figure, 
the dotted curve is the result of the Hartree approximation, where 
the proton EFF is the same as in free space. 
The dashed curve is the result of the full 
RRPA, without the modifications 1 and 2. The dot-dashed curve shows the
result of the full RRPA with the meson-N form factor but $R_{pE}=1$. 
The upper (lower) solid curve shows the result of the full RRPA for
$m_q = 5~(300)$ MeV, including all modifications.   
Because of the density dependent coupling,
the reduction of the response function from the Hartree
result, caused by the full RRPA, becomes much smaller than that in QHD-I.
On the other hand, 
the modification of the proton EFF is very
significant, yielding a much bigger reduction in the response.
We can see that the effect of the meson-N form factor 
is relatively minor. 

It is also interesting to see the quark mass dependence of the LR.
As an example, we consider the case of $m_q =
300$ MeV.  In comparison with the case $m_q$ = 5 MeV,
it is a little smaller and the peak position is shifted to the higher
energy transfer side.  This is related to the smaller effective nucleon
mass in the case $m_q$ = 300 MeV than when $m_q$ = 5 MeV.

The Coulomb sum, $C(q) = \int^q_0 dq_0 \/ S_L(q, q_0)$, 
is shown in  Fig. \ref{fig:csr} as a function of three-momentum transfer,
$q$.  For high $q$, the strength is about 20\% lower in the full
calculation (the solid and dot-dashed curves are
for $m_q$ = 5 and 300 MeV, respectively)  
than for the Hartree response with $R_{pE}=1$ (the dotted curve).  
For low $q$, the full
calculation with the constituent quark mass remains much lower
than the Hartree result, while in case of the light quark mass it
gradually approaches the Hartree one.  This difference is caused by that
the effective nucleon mass for $m_q$ = 5 MeV being larger in matter than that
for $m_q$ = 300 MeV.

We comment on the transverse response (TR) from nuclear
matter.  In Ref. \cite{emff} it was found that
the in-medium modification of the nucleon magnetic form factor within QMC
is very small.  Therefore,
one would expect the total change in the TR caused by
RRPA correlations and the effect of the variation of the structure of
the nucleon to be much smaller than in the LR.

\section{Summary} 

We have calculated the LR of nuclear
matter using the QMC model.  The reduction of the
$\sigma$-N coupling constant with density decreases the
contribution of the RRPA, while the modification of the proton EFF  
in medium reduces the LR considerably.  The LR or the Coulomb sum is
reduced by about 20\% in total, with RRPA correlations and the variation
of the in-medium nucleon structure contributing about fifty-fifty.
It will be interesting to extend this work to calculate both the
LR and TR functions for finite nuclei,
in order to compare directly with the new experimental
results which are anticipated soon \cite{new}. 

The author would like to thank A.W. Thomas, K. Tsushima, M. Ericson,
P.A.M. Guichon, W. Bentz and J. Morgenstern for valuable discussions.
This work was supported by the Japan Society for the Promotion of Science.

\end{document}